\documentclass[letter]{aa}  

\usepackage{graphicx}
\usepackage{txfonts}
\usepackage{lipsum}
\usepackage{subcaption}         
\usepackage{lscape}             
\usepackage{placeins}           
                                
\usepackage{adjustbox}

\usepackage[]{hyperref}
\hypersetup{unicode=true, colorlinks=true, linkcolor=[rgb]{0.53, 0.15, 0.34}, citecolor=[rgb]{0.0, 0.27, 0.42}, filecolor=[rgb]{1.0, 0.13, 0.32}, urlcolor=[rgb]{0.53, 0.15, 0.34}}

\usepackage{amsmath}
\usepackage[mathscr]{euscript} 
\usepackage{breqn} 
\usepackage{comment}
                            
\begin{document}

   \title{The unidentified PeV source LHAASO J2108+5157: \\ A microquasar remnant?}


%
%
%

\author{Josep Mart\'{\i}\inst{1}\corrauth{jmarti@ujaen.es}
\and Pedro L. Luque-Escamilla\inst{2}\email{peter@ujaen.es}
\and Leandro Abaroa\inst{3,4}\email{leandroabaroa@gmail.com}
\and Gustavo E. Romero\inst{3,4}\email{gustavo.esteban.romero@gmail.com}
\and Valent\'{\i} Bosch-Ramon\inst{5}\email{vbosch@fqa.ub.edu}
\and ~~~~ Josep M. Paredes\inst{5}\email{jmparedes@ub.edu}
\and Cintia S. Peri\inst{1}\email{csperi@ujaen.es}
}

\institute{Departamento de F\'isica. EPS Ja\'en. Universidad de Ja\'en. Campus Las Lagunillas s/n 23071 Spain
\and Departamento de Ingenier\'ia Mec\'anica y Minera, EPS Ja\'en. Universidad de Ja\'en. Campus Las Lagunillas s/n 23071 Spain
\and Instituto Argentino de Radioastronomía (CCT La Plata, CONICET; CICPBA; UNLP), C.C.5, (1894), Villa Elisa, Argentina
\and Facultad de Ciencias Astron\'omicas y Geof\'{\i}sicas, Universidad Nacional de La Plata, B1900FWA La Plata, Argentina
\and Departament de F\'isica Qu\`antica i Astrof\'isica, Institut de Ci\`encies del Cosmos, Universitat de Barcelona, IEEC-UB, Mart\'i i Franqu\`es 1, 08028 Barcelona, Spain
}

   \date{Received May XX, 2026}

 
  \abstract
  {LHAASO~J2108+5157 is one of the most intriguing Galactic ultra-high-energy $\gamma$-ray sources. It is detected up to hundreds of TeV but lacks a clear counterpart at lower frequencies. \textit{Fermi}-LAT observations have revealed a compact GeV source spatially coincident with the PeV emission, while VERITAS and LST-1 impose stringent upper limits in the intermediate TeV band, producing a pronounced spectral suppression between the GeV and TeV domains.}
  {We investigate a time-dependent hadronic scenario in which a past episode of microquasar jet activity injected relativistic protons that are now observed in a relic phase.}
  {Energy-dependent diffusion naturally segregates the particle population: Lower-energy protons remain confined near dense molecular gas and account for the GeV emission, whereas higher-energy particles reach more distant material, producing the hard TeV--PeV component.}
{For representative interstellar conditions, the model is simultaneously consistent with the broadband spectral energy distribution, the spatial compact--extended dichotomy, and the lack of detectable radio or X-ray counterparts.}
  {The observed bimodal $\gamma$-ray phenomenology thus emerges as a transport effect of a single injected proton population, supporting the interpretation of LHAASO~J2108+5157 as a fossil microquasar remnant observed at an intermediate evolutionary stage and highlighting relic jet systems as a viable class of Galactic PeVatrons.}

   
   \maketitle
   
   \nolinenumbers


\section{Introduction}
\label{sec:intro}

The origin of Galactic cosmic rays up to PeV energies remains one of the central open problems in high-energy astrophysics. The recent discovery of numerous ultra-high-energy $\gamma$-ray sources by the Large High Altitude Air Shower Observatory (LHAASO) has revealed the presence of powerful Galactic accelerators capable of reaching PeV energies, commonly referred to as PeVatrons \citep[e.g.][]{Cao2024_Catalog}. However, identifying the nature of these accelerators is challenging, particularly for sources lacking clear counterparts at other wavelengths.

LHAASO~J2108+5157 is one of the most enigmatic members of this population. The source has been firmly detected at TeV to PeV energies by both LHAASO \citep[][]{Cao2021_Discovery} and the High-Altitude Water Cherenkov Gamma-Ray Observatory 
(HAWC) \citep{Adams_2025}, yet searches at lower energies fail to reveal an obvious accelerator. \emph{Fermi}-LAT observations have revealed a compact GeV $\gamma$-ray source spatially coincident with the LHAASO position \citep{Abe2023_Multiwavelength, Adams_2025}.
In the energy range bridging GeV and TeV bands, imaging atmospheric Cherenkov telescopes, including VERITAS and the Large-Sized Telescope (LST-1) of CTAO, provide stringent upper limits, indicating a pronounced spectral suppression \citep{Abe2023_Multiwavelength, Adams_2025}.

Several astrophysical scenarios have been proposed to explain the 
nature of
 LHAASO~J2108+5157, ranging from a supernova remnant (SNR) interacting with molecular clouds \citep{DeSarkar2023_SNRModel, Mitchell2024_SNRCloud} to a pulsar wind nebula or TeV halo powering the
high-energy emission \citep{Cao2021_Discovery, Abe2023_Multiwavelength, Adams_2025}. However, deep multi-wavelength searches have revealed no obvious counterpart, and recent studies have firmly ruled out 
an active microquasar (MQ) in the vicinity \citep{Marti2026_LHAASO}. The continued absence of counterparts across the electromagnetic spectrum challenges associations with standard active accelerators, leaving LHAASO~J2108+5157 as an enigmatic `dark' accelerator.

A notable feature of LHAASO J2108+5157 is the apparent spectral separation between the GeV and TeV--PeV emission. The GeV component detected by \textit{Fermi}-LAT is steady and compatible with a point source within the resolution of the instrument, while the TeV--PeV emission observed by HAWC and LHAASO appears point-like or mildly extended.
Moreover, the spectral energy distribution (SED) exhibits a gap between GeV and TeV energies constrained by Cherenkov telescope limits. These profound spatial and spectral differences have led to debates regarding their physical association \citep{Abdollahi2020, Mitchell2024_SNRCloud}, as they strongly challenge standard interpretations based on a single steady accelerator.

An alternative possibility is that the source represents the relic
phase of a powerful MQ jet. During their active stage,
MQs can inject substantial amounts of relativistic particles
into the surrounding interstellar medium (ISM). Once the central engine
switches off, the injected particles continue to propagate and
interact with nearby molecular material, potentially producing
long-lived $\gamma$-ray emission via $pp$ collisions.
During this phase, any extended synchrotron emission from primary electrons typically falls below detection limits \citep{2026arXiv260817000A}.

In this work we investigate this hypothesis using the
time-dependent MQ remnant (MQR) framework developed by
\citet{Abaroa_etal_2026MQR}. In this model, a single proton
population injected during the jet phase evolves 
into two observationally distinct $\gamma$-ray emitting particle populations
as a result of energy-dependent diffusion, confinement, and interaction with the local ISM (ISM; see also \citealt{valenti2005} for a similar scenario for active MQs).
We show that within plausible interstellar conditions, this
scenario provides a viable existence-proof model that 
can simultaneously account for the broadband SED, the spatial compact--extended dichotomy,
and the apparent spectral separation between the GeV and
TeV--PeV components without invoking multiple acceleration
episodes.

\section{Observational overview}
\label{sec:data}

\subsection{Gamma-ray data}

LHAASO J2108+5157 has been firmly detected by LHAASO at energies exceeding 100 TeV \citep[][]{Cao2021_Discovery}, with HAWC confirming a spectrum extending smoothly into the multi-TeV domain \citep{Adams_2025}. This very-high-energy emission is consistent with a point-like or mildly extended source. At lower energies, \textit{Fermi}-LAT has revealed a steady, compact GeV source spatially coincident with the LHAASO position. Its spectrum is best described by a log-parabola peaking at sub-GeV energies and steepening rapidly above a few GeV \citep{Abe2023_Multiwavelength}. Crucially, VERITAS and LST-1 provide stringent upper limits around $\sim 1$~TeV, creating a pronounced spectral `valley' between the GeV and TeV--PeV bands \citep{Adams_2025}. This challenges simple single-zone interpretations.

In this work we adopt the dedicated \textit{Fermi}-LAT analysis of \citet{Abe2023_Multiwavelength}.
This analysis is restricted to energies above $\sim1$ GeV, thereby minimising nearby pulsar contamination.

\subsection{Multi-wavelength constraints}

Extensive searches have revealed no obvious accelerator associated with LHAASO J2108+5157. The region lacks detectable non-thermal radio or X-ray emission, thus strongly disfavouring simple  bright pulsar wind nebulae or active SNRs \citep{Abe2023_Multiwavelength, Adams_2025}. A previously proposed MQ candidate \citep{Mahanta2024_uGMRT} has recently been re-identified as a background extragalactic object, leaving no active Galactic candidates detected in the field so far \citep{Marti2026_LHAASO}. However, 
CO and HI observations have revealed substantial molecular material in the direction of the source. Throughout this work we adopt MML[2017]4607 ($d\simeq3.3$ kpc, $n\sim30$ cm$^{-3}$; \citealt{Miville2017, deLaFuente2023_MC}), following its original association with LHAASO J2108+5157 \citep{Cao2021_Discovery}, as the representative hadronic target for our transport model. 
Subsequent higher-resolution studies resolved this emission into several molecular components and discussed alternative distances \citep{deLaFuente2023_MC2}. However, because these alternative estimates are strongly influenced by trigonometric-parallax priors and trace denser substructures not fully coincident with the $\gamma$-ray peak, we retained the kinematic distance estimate of $d\simeq 3.3$~kpc and the fiducial density $n\sim30$~cm$^{-3}$ as representative of the interaction region.

\subsection{Phenomenological spectral decomposition}
\label{sec:phenomenological_sed}

The broadband SED is well reproduced by the sum of two power laws with exponential cut-offs (ECPL), 
which naturally captures the pronounced spectral suppression between the GeV and TeV bands. This purely phenomenological decomposition reveals a soft low-energy component ($\Gamma_1 = 2.38 \pm 0.51$, $E_{\mathrm{cut},1} = 43 \pm 15~\mathrm{GeV}$) consistent with a diffusion-relaxed population and a harder high-energy component ($\Gamma_2 = 1.65 \pm 0.22$, $E_{\mathrm{cut},2} = 635 \pm 180~\mathrm{TeV}$) extending to PeV energies with significantly less spectral softening. Because this description uses a dedicated LAT dataset and a two-component model, its fitted parameters naturally differ from the single-component 
 ECPL fit in \citet{Adams_2025}. Assuming standard hadronic interactions ($n \approx 30\ \mathrm{cm}^{-3}$ at $d = 3.3~\mathrm{kpc}$), the required proton energy budget is modest ($W_{p,\mathrm{tot}} \sim 2.9 \times 10^{48}~\mathrm{erg}$). The pronounced spectral hardening ($\Delta\Gamma \approx 0.73$) strongly suggests spatially differentiated populations driven by energy-dependent diffusion from a single accelerator, motivating the physical transport model developed in Sect.~\ref{sec:model}.

\section{Physical model}
\label{sec:model}

We followed  \citet{Abaroa_etal_2026MQR} to model the system.
The power required for protons to reach PeV energies in a single accelerator is $L_{\rm K}\gtrsim 10^{39}(E_{\rm max,p}/10\,{\rm PeV})^{2}\,{\rm erg\,s^{-1}}$ \citep{Wang2025ApJ}.
Therefore,
we considered a high-mass MQ with a jet kinetic power of
$L_j = 3\times10^{39}$ erg s$^{-1}$ active for
$t_{\rm MQ}=10$ kyr followed by a relic phase lasting 
$t_{\rm MQR}\simeq 3$ kyr, leading to a global age of the system of $t_{\rm total} \simeq 13$ kyr. 

Particles were injected with a power-law spectrum consistent with
diffusive shock acceleration.
A fraction, $q_{\rm rel}=0.1$, of the jet power
was converted into non-thermal particles, with a hadron-to-lepton power
ratio of $100$, which means that 99\% of the available power resides in protons \citep{Abaroa_etal2024(S26)}. 
This required a physically plausible total hadronic energy budget of $\sim 10^{49}$--$10^{50}$~erg over the active phase. 
If the ambient density is $\sim 1.67\times 10^{-25}\,{\rm g\,cm^{-3}}$ \citep{Spitzer-Book1978}, the cocoon had a semi-major axis of $l_{\rm c}\sim 20\,$pc at $t_0$ \citep{1997MNRAS.286..215K,Abaroa_etal2024(S26)}. To have dynamics similar to that of a radio galaxy, the semi-minor axis should have been $w_{\rm c}\approx l_{\rm c}/3\approx 7\,$pc (e.g. \citealt{2009A&A...497..325B}). The diluted cavity has a number density of $10^{-3}\,{\rm cm^{-3}}$ and a typical magnetic field of $3\,{\mu\rm G}$. We assumed that these two parameters remain roughly constant throughout the life of the MQR. 
Within the cocoon, we assumed a standard turbulent medium with a Kolmogorov spectrum (see the appendix in \citealt{Abaroa_etal_2026MQR}).

During the active phase, the jet
injects relativistic particles into the cocoon. After jet switch-off, 
the particles 
continue to escape from the relic cocoon and propagate
through the ISM.

Once protons escaped from the cocoon, their distribution in the ISM,
$n_{\rm I}(E,R,t)$, was evolved with the spherically symmetric diffusion
equation,
\begin{equation}
\frac{\partial n_{\rm I}}{\partial t}
=
D(E)\frac{1}{R^2}\frac{\partial}{\partial R}
\left(R^2\frac{\partial n_{\rm I}}{\partial R}\right)
+
Q_{\rm esc}(E,t)\,\delta^3(R)
\label{eq:diffISM}
,\end{equation}
where continuous energy losses during propagation can be neglected over the timescales considered. 
The escape rate from the cocoon was estimated as $Q_{\rm esc}(E,t)=
n_{\rm c}(E,t)V_{\rm c}(t)/t_{\rm esc,c}(E,t)$ \citep{Abaroa_etal_2026MQR},
with $n_{\rm c}$ as the cocoon proton density, $V_{\rm c}$ as the cocoon
volume, and $t_{\rm esc,c}$ as the energy-dependent escape time. The
solution at the position of each effective interaction region was obtained by
convolving this escape history with the Green function for diffusion
\citep{Aharonian&Atoyan_1996},
\begin{equation}
n_{\rm I}(E,R,t)=
\int_0^t
\frac{Q_{\rm esc}(E,t')\,
\exp[-R^2/4D(E)(t-t')]}
{[4\pi D(E)(t-t')]^{3/2}}
\,{\rm d}t' .
\label{eq:green_solution}
\end{equation}
Thus, the spectra at the two interaction regions were evaluated at the same global
system age, but each target sampled a different part of the propagated
proton distribution.

The ISM diffusion coefficient was parametrized as
\begin{equation}
D(E)=D_0\left(\frac{E}{10\,{\rm GeV}}\right)^\delta ,
\label{eq:diffcoef}
\end{equation}
where we assumed $D_0=10^{26}\,{\rm cm^2\,s^{-1}}$ and $\delta=0.5$. This
corresponds to a suppression factor of $\sim10^{-2}$ relative to the
average Galactic value at 10 GeV, as expected in turbulent environments
around compact accelerators \citep[e.g.][]{Aharonian&Atoyan_1996,Gabici+2007,Fujita_2010, NavaGabici2012}.

The $\gamma$-ray emission from each interaction region was computed through
inelastic $pp$ interactions followed by $\pi^0$ decay. For a homogeneous
target of density $n_{\rm H}$, we used a delta-functional approximation
for the neutral-pion source function,
\begin{equation}
q_{\pi^0}(E_\pi,R,t)=
\frac{c\,n_{\rm H}}{\kappa_\pi}
\sigma_{pp}(E_p)\,
n_{\rm I}(E_p,R,t),
\label{eq:pion_source}
\end{equation}
where $\kappa_\pi=0.17$ and
$E_\pi=\kappa_\pi(E_p-m_pc^2)$. The inelastic cross-section was taken
from the parametrisation of \citet{Kelner2006}. The photon
emissivity from $\pi^0\rightarrow2\gamma$ was then
\begin{equation}
q_\gamma(E_\gamma,R,t)
=
2
\int_{E_{\pi,\min}}^\infty
\frac{q_{\pi^0}(E_\pi,R,t)}
{\sqrt{E_\pi^2-m_{\pi^0}^2c^4}}
\,{\rm d}E_\pi ,
\label{eq:gamma_emissivity}
\end{equation}
with
$E_{\pi,\min}=E_\gamma+m_{\pi^0}^2c^4/(4E_\gamma)$.
The SED of each interaction target was obtained by multiplying $q_\gamma$ by the
corresponding volume. The
main parameters of our model are listed in Table~\ref{tab:parametros generales}.

\section{Results and discussion}
\label{sec:results_discussion}

We solved the full time-dependent transport equations detailed in Sect.~\ref{sec:model} following the framework of
\citet{Abaroa_etal_2026MQR} for a system with an active phase of
$t_{\rm MQ}=10$ kyr followed by a relic phase up to 5 kyr.
As discussed below, we find that a current relic age of $t_{\rm MQR} \sim 3$ kyr provides the best description of the data.

\begin{figure}
    \centering
    \includegraphics[width=8.38cm]{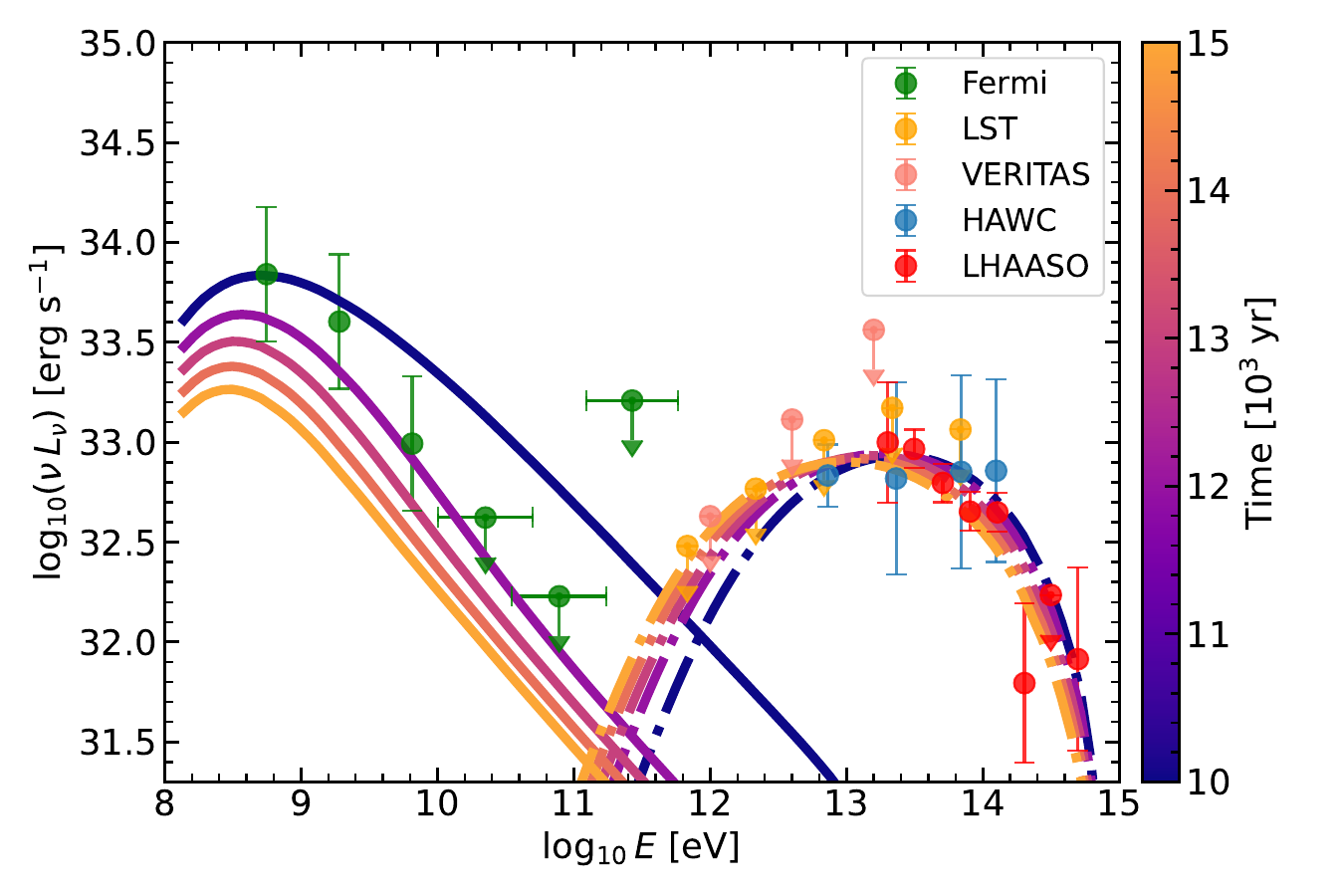}
    \caption{SEDs of the $\gamma$-ray emission from both interaction regions. Solid lines (with a peak luminosity at GeV energies) correspond to $C_1$, while dotted-dashed lines (with a peak luminosity at TeV energies) correspond to $C_2$. Different colours indicate model spectra evaluated at different evolutionary times, as specified in the text, not distinct particle populations. The assumed distance to the MQR is $d=3.3\,$kpc. The current epoch is best represented by $t\sim 13\,$kyr, which corresponds to the violet lines.}
    \label{fig:sed}
\end{figure}

\begin{figure}
    \centering
    \includegraphics[width=8.38cm]{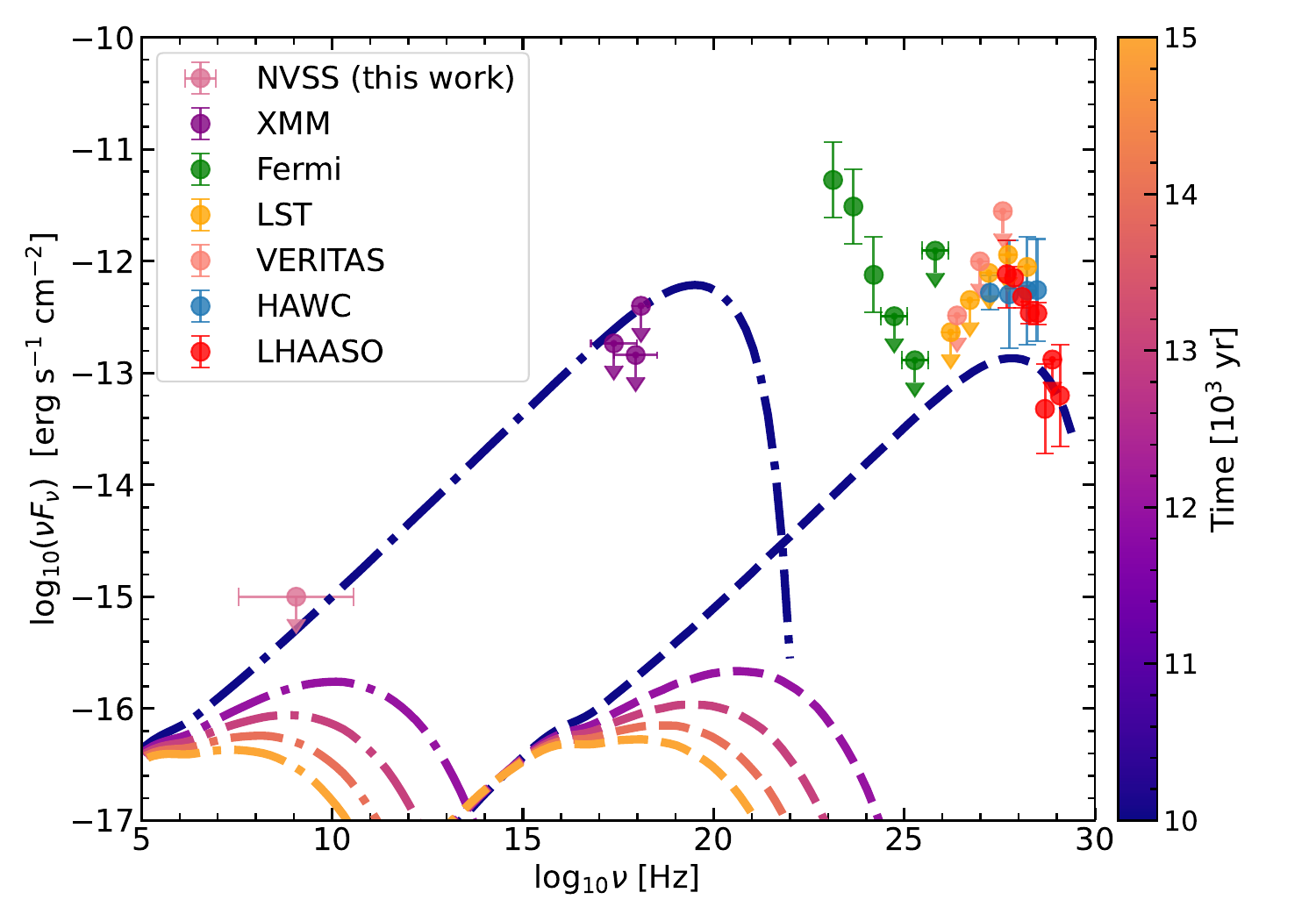}
    \caption{Flux of the cocoon. The emission is produced by relativistic electrons when they interact with magnetic (synchrotron, dashed-dotted lines) and radiation (inverse Compton, dashed lines) fields. The colour-scale bar gives the total system age, $t_{\rm total}$. The current epoch is best represented by $t\sim 13\,$kyr, which corresponds to the violet lines. The $\gamma$-ray points are not associated with the cocoon emission and are shown only for comparison.} 
    \label{fig: cocoon_flux}
\end{figure}

The target material for the cosmic rays is represented by two effective
interaction regions within the MML[2017]4607 molecular complex at
$d=3.3$ kpc; both have a characteristic density of
$n\simeq30\,{\rm cm^{-3}}$. These regions should not be interpreted
as two resolved gamma-ray centroids but as a simplified description
of the dense gas sampled by different parts of the propagated proton
population. The inner region, $C_1$, represents a localised region of dense material in direct contact with the remnant, and $R_1$ is an effective propagation scale at the centre of such a region for the lower-energy particles responsible for the GeV emission. The second region, $C_2$, is a more distant effective target located at
$R_2\simeq35$ pc, with $r_2=12$ pc, and it accounts for the interaction of
the fast-escaping PeVatron component.

At the adopted distance, the separation between the two effective
interaction regions corresponds to an angular scale of $\sim0.6^\circ$.
In the present model, this should be regarded as a characteristic scale
within the molecular complex, not as a prediction of two clearly resolved
gamma-ray sources. The scenario therefore predicts an energy-dependent
morphology, with the GeV and TeV--PeV emission tracing different parts
of the same target complex. 
Observationally, any underlying centroid shift remains unresolved due to current instrumental limits. For instance, the broad $\sim 0.8^\circ$ \textit{Fermi}-LAT point spread function at 1~GeV easily encompasses both emission regions, and projection effects along the line of sight can further reduce the apparent angular separation.
A quantitative morphological fit is beyond
the scope of this work
and will require 
energy-resolved
$\gamma$-ray maps with a greater number of statistics.

Figure~\ref{fig:sed} shows the SEDs produced via $pp$ interactions in each interaction region at different
evolutionary stages. Emission from $C_1$ peaks at GeV energies,
while $C_2$ dominates in the TeV--PeV range. At the current epoch (violet line, corresponding to $t_{\rm total} \simeq 13\,$kyr),
the model simultaneously reproduces (i) the compact GeV component
detected by \textit{Fermi}-LAT, arising from accumulated lower-energy
protons in $C_1$; (ii) the hard TeV--PeV emission observed by HAWC
and LHAASO, produced by the high-energy protons
interacting in $C_2$; and (iii) the pronounced suppression between the
GeV and TeV bands constrained by VERITAS and LST-1.
This spectral `valley' emerges naturally from transport:
Intermediate-energy protons diffuse beyond the densest gas before
accumulating a significant $pp$ emissivity, 
reducing their $\gamma$-ray
yield in the $\sim0.1$--$1$ TeV domain. 
The apparent bimodality of the SED therefore reflects the evolutionary stage of the system and the spatial distribution of the target material.
Furthermore, the compact nature of the \textit{Fermi}-LAT source is fully compatible with our scenario, as the low-energy protons responsible for the GeV emission have a significantly shorter diffusion length and remain confined near the inner region.

 The simultaneous presence of confined
GeV emission and escaping PeV emission constrains the system to an
intermediate evolutionary stage. This means it must be old enough for PeV particles
to have accumulated in a distant molecular material yet young enough for their
density to not have been completely diluted into the ISM.

To assess the detectability of the relic structure itself, we computed the
evolution of the primary electron population inside the cocoon and its
associated synchrotron and inverse Compton emission when interacting with the magnetic and radiation fields.
Figure~\ref{fig: cocoon_flux} shows the predicted broadband cocoon
flux as a function of time. At the present epoch, the synchrotron
component lies well below current radio and X-ray sensitivities,
consistent with the non-detection of any clear low-energy counterpart
and with the low surface brightness expected for MQR
cocoons \citep{2026arXiv260817000A}.
Primary electrons cool and dilute rapidly during the relic phase,
whereas long-lived protons continue to diffuse into the surrounding
molecular complex and power the observed $\gamma$-ray emission.
The system thus behaves as a passive hadronic injector embedded in
dense gas.

Alternative interpretations face significant challenges. Leptonic
scenarios, such as relic pulsar wind nebulae or TeV halos \citep{Abe2023_Multiwavelength, Adams_2025}
require magnetic fields
well below typical Galactic values ($B\ll 3\,\mu$G) to suppress
synchrotron emission while maintaining ultra-high-energy inverse Compton
fluxes. 
Similarly, hadronic scenarios in which a young SNR
accelerates protons that later interact with nearby molecular
material \citep{Mitchell2024_SNRCloud} face strong observational constraints. An SNR capable of
supplying multi-TeV to PeV particles would generally be expected
to exhibit detectable non-thermal radio and/or X-ray emission.
The absence of a clear remnant signature in current surveys
therefore disfavours such interpretations.
By contrast, the MQR
scenario naturally anticipates a multi-wavelength--quiet state.

The total energy in relativistic protons required to reproduce the
observed $\gamma$-ray luminosity, estimated from the phenomenological
hadronic fit with
 $n_{\rm gas}\sim 30$ cm$^{-3}$, is
$W_p \approx 2.9\times10^{48}$ erg. This value is comfortably
below the total hadronic energy that can be supplied by a jet with
$L_{\rm j}\sim10^{39}$ erg s$^{-1}$ over $10^4$ yr for a reasonable
fraction of power channelled into relativistic protons.
The transport model thus 
operates within energetically
plausible parameters rather than requiring extreme assumptions.

If our MQR origin is correct, LHAASO~J2108+5157 may represent
a prototype of a previously overlooked class of Galactic PeVatrons:
fossil MQ jets whose engines have already switched off.
Energy-dependent transport in relic jet environments may therefore
play a significant role in shaping the observable high-energy sky and
in contributing to the Galactic cosmic-ray population at PeV energies.

\section{Conclusions}
\label{sec:conclusions}

We have investigated the ultra-high-energy $\gamma$-ray source LHAASO~J2108+5157 within a time-dependent MQR framework and concluded that its apparently disconnected GeV and TeV--PeV components can be explained by a single relativistic proton population injected during a past MQ phase. Energy-dependent diffusion naturally drives this spatial and spectral segregation, as lower-energy protons remain confined near the former jet termination region, producing the compact GeV emission, whereas escaping higher-energy protons reach distant molecular material to generate the hard PeV component. Consequently, the pronounced GeV--TeV spectral suppression can arise
as a transport effect instead of requiring distinct sources. 
The simultaneous detection of these confined and escaping populations 
is naturally accommodated by a system at 
an intermediate evolutionary stage, 
such as 
 a jet that switched off $\sim 3$~kyr ago following a $\sim 10$~kyr active phase. This fossil scenario naturally explains the absence of non-thermal X-ray and radio counterparts, as primary electrons cool and dilute rapidly while long-lived protons continue powering hadronic interactions. Ultimately, LHAASO~J2108+5157 may not be a classical young SNR or a pulsar-driven PeVatron but instead the prototype of a hidden population of extinct MQs, thus highlighting the crucial role of energy-dependent transport in relic jet environments in shaping the Galactic PeV cosmic-ray landscape.

\begin{acknowledgements}
The authors acknowledge support from projects 
PID2022-136828NB-C41, PID2022-136828NB-C42,
PID2025-168247NB-C41 and PID2025-168247NB-C42
funded by the Spanish MCIN/AEI/10.13039/501100011033 and ``ERDF A way of making Europe'', and
through the Unit of Excellence María de Maeztu
awards to the ICCUB (CEX2024-
001451-M).
We also acknowledge support from 
Plan Andaluz de Investigaci\'on, Desarrollo e Innovaci\'on  as research group FQM-322. 
VB-R is Correspondent Researcher of CONICET, Argentina, at the IAR.
\end{acknowledgements}

%
\bibliographystyle{aa} 
 \bibliography{LHAASO_Jet_Relic} 

\newcommand{\noopsort}[1]{}
\begin{thebibliography}{26}
\expandafter\ifx\csname natexlab\endcsname\relax\def\natexlab#1{#1}\fi

\bibitem[{{Abaroa} {et~al.}(2026{\natexlab{a}}){Abaroa}, {Romero}, \&
  {Bosch-Ramon}}]{2026arXiv260817000A}
{Abaroa}, L., {Romero}, G.~E., \& {Bosch-Ramon}, V. 2026{\natexlab{a}}, arXiv
  e-prints, arXiv:2608.17000

\bibitem[{{Abaroa} {et~al.}(2026{\natexlab{b}}){Abaroa}, {Romero}, \&
  {Bosch-Ramon}}]{Abaroa_etal_2026MQR}
{Abaroa}, L., {Romero}, G.~E., \& {Bosch-Ramon}, V. 2026{\natexlab{b}}, \aap,
  705, L4

\bibitem[{{Abaroa} {et~al.}(2024){Abaroa}, {Romero}, {Mancuso}, \&
  {Rizzo}}]{Abaroa_etal2024(S26)}
{Abaroa}, L., {Romero}, G.~E., {Mancuso}, G.~C., \& {Rizzo}, F.~N. 2024, \aap,
  691, A93

\bibitem[{{Abdollahi} {et~al.}(2020){Abdollahi}, {Acero}, {Ackermann},
  {Ajello}, {Atwood}, {Axelsson}, {Baldini}, {Ballet}, {Barbiellini},
  {Bastieri}, {Becerra Gonzalez}, {Bellazzini}, {Berretta}, {Bissaldi},
  {Blandford}, {Bloom}, {Bonino}, {Bottacini}, {Brandt}, {Bregeon}, {Bruel},
  {Buehler}, {Burnett}, {Buson}, {Cameron}, {Caputo}, {Caraveo}, {Casandjian},
  {Castro}, {Cavazzuti}, {Charles}, {Chaty}, {Chen}, {Cheung}, {Chiaro},
  {Ciprini}, {Cohen-Tanugi}, {Cominsky}, {Coronado-Bl{\'a}zquez}, {Costantin},
  {Cuoco}, {Cutini}, {D'Ammando}, {DeKlotz}, {de la Torre Luque}, {de Palma},
  {Desai}, {Digel}, {Di Lalla}, {Di Mauro}, {Di Venere}, {Dom{\'\i}nguez},
  {Dumora}, {Fana Dirirsa}, {Fegan}, {Ferrara}, {Franckowiak}, {Fukazawa},
  {Funk}, {Fusco}, {Gargano}, {Gasparrini}, {Giglietto}, {Giommi}, {Giordano},
  {Giroletti}, {Glanzman}, {Green}, {Grenier}, {Griffin}, {Grondin}, {Grove},
  {Guiriec}, {Harding}, {Hayashi}, {Hays}, {Hewitt}, {Horan},
  {J{\'o}hannesson}, {Johnson}, {Kamae}, {Kerr}, {Kocevski}, {Kovac'evic'},
  {Kuss}, {Landriu}, {Larsson}, {Latronico}, {Lemoine-Goumard}, {Li},
  {Liodakis}, {Longo}, {Loparco}, {Lott}, {Lovellette}, {Lubrano}, {Madejski},
  {Maldera}, {Malyshev}, {Manfreda}, {Marchesini}, {Marcotulli},
  {Mart{\'\i}-Devesa}, {Martin}, {Massaro}, {Mazziotta}, {McEnery}, {Mereu},
  {Meyer}, {Michelson}, {Mirabal}, {Mizuno}, {Monzani}, {Morselli},
  {Moskalenko}, {Negro}, {Nuss}, {Ojha}, {Omodei}, {Orienti}, {Orlando},
  {Ormes}, {Palatiello}, {Paliya}, {Paneque}, {Pei}, {Pe{\~n}a-Herazo},
  {Perkins}, {Persic}, {Pesce-Rollins}, {Petrosian}, {Petrov}, {Piron}, {Poon},
  {Porter}, {Principe}, {Rain{\`o}}, {Rando}, {Razzano}, {Razzaque}, {Reimer},
  {Reimer}, {Remy}, {Reposeur}, {Romani}, {Saz Parkinson}, {Schinzel},
  {Serini}, {Sgr{\`o}}, {Siskind}, {Smith}, {Spandre}, {Spinelli}, {Strong},
  {Suson}, {Tajima}, {Takahashi}, {Tak}, {Thayer}, {Thompson}, {Tibaldo},
  {Torres}, {Torresi}, {Valverde}, {Van Klaveren}, {van Zyl}, {Wood},
  {Yassine}, \& {Zaharijas}}]{Abdollahi2020}
{Abdollahi}, S., {Acero}, F., {Ackermann}, M., {et~al.} 2020, \apjs, 247, 33

\bibitem[{{Abe} {et~al.}(2023){Abe}, {Aguasca-Cabot, A.}, {Agudo, I.}, {Alvarez
  Crespo, N.}, {Antonelli, L. A.}, {Aramo, C.}, {Arbet-Engels, A.}, {Artero,
  M.}, {Asano, K.}, {Aubert, P.}, {Baktash, A.}, {Bamba, A.}, {Baquero Larriva,
  A.}, {Baroncelli, L.}, {Barres de Almeida, U.}, {Barrio, J. A.}, {Batkovic,
  I.}, {Baxter, J.}, {Becerra González, J.}, {Bernardini, E.}, {Bernardos, M.
  I.}, {Bernete Medrano, J.}, {Berti, A.}, {Bhattacharjee, P.}, {Biederbeck,
  N.}, {Bigongiari, C.}, {Bissaldi, E.}, {Blanch, O.}, {Bordas, P.}, {Buisson,
  C.}, {Bulgarelli, A.}, {Burelli, I.}, {Buscemi, M.}, {Cardillo, M.}, {Caroff,
  S.}, {Carosi, A.}, {Cassol, F.}, {Cauz, D.}, {Ceribella, G.}, {Chai, Y.},
  {Cheng, K.}, {Chiavassa, A.}, {Chikawa, M.}, {Chytka, L.}, {Cifuentes, A.},
  {Contreras, J. L.}, {Cortina, J.}, {Costantini, H.}, {D’Amico, G.},
  {Dalchenko, M.}, {De Angelis, A.}, {de Bony de Lavergne, M.}, {De Lotto, B.},
  {de Menezes, R.}, {Deleglise, G.}, {Delgado, C.}, {Delgado Mengual, J.},
  {della Volpe, D.}, {Dellaiera, M.}, {Di Piano, A.}, {Di Pierro, F.}, {Di
  Tria, R.}, {Di Venere, L.}, {Díaz, C.}, {Dominik, R. M.}, {Dominis Prester,
  D.}, {Donini, A.}, {Dorner, D.}, {Doro, M.}, {Elsässer, D.}, {Emery, G.},
  {Escudero, J.}, {Fallah Ramazani, V.}, {Ferrara, G.}, {Fiasson, A.}, {Freixas
  Coromina, L.}, {Fröse, S.}, {Fukami, S.}, {Fukazawa, Y.}, {Garcia, E.},
  {Garcia López, R.}, {Gasparrini, D.}, {Geyer, D.}, {Giesbrecht Paiva, J.},
  {Giglietto, N.}, {Giordano, F.}, {Giro, E.}, {Gliwny, P.}, {Godinovic, N.},
  {Grau, R.}, {Green, D.}, {Green, J.}, {Gunji, S.}, {Hackfeld, J.}, {Hadasch,
  D.}, {Hahn, A.}, {Hashiyama, K.}, {Hassan, T.}, {Hayashi, K.}, {Heckmann,
  L.}, {Heller, M.}, {Herrera Llorente, J.}, {Hirotani, K.}, {Hoffmann, D.},
  {Horns, D.}, {Houles, J.}, {Hrabovsky, M.}, {Hrupec, D.}, {Hui, D.},
  {Hütten, M.}, {Imazawa, R.}, {Inada, T.}, {Inome, Y.}, {Ioka, K.}, {Iori,
  M.}, {Ishio, K.}, {Iwamura, Y.}, {Jacquemont, M.}, {Jimenez Martinez, I.},
  {Jurysek, J.}, {Kagaya, M.}, {Karas, V.}, {Katagiri, H.}, {Kataoka, J.},
  {Kerszberg, D.}, {Kobayashi, Y.}, {Kong, A.}, {Kubo, H.}, {Kushida, J.},
  {Lainez, M.}, {Lamanna, G.}, {Lamastra, A.}, {Le Flour, T.}, {Linhoff, M.},
  {Longo, F.}, {López-Coto, R.}, {López-Moya, M.}, {López-Oramas, A.},
  {Loporchio, S.}, {Lorini, A.}, {Luque-Escamilla, P. L.}, {Majumdar, P.},
  {Makariev, M.}, {Mandat, D.}, {Manganaro, M.}, {Manicò, G.}, {Mannheim, K.},
  {Mariotti, M.}, {Marquez, P.}, {Marsella, G.}, {Martí, J.}, {Martinez, O.},
  {Martínez, G.}, {Martínez, M.}, {Marusevec, P.}, {Mas-Aguilar, A.},
  {Maurin, G.}, {Mazin, D.}, {Mestre Guillen, E.}, {Micanovic, S.}, {Miceli,
  D.}, {Miener, T.}, {Miranda, J. M.}, {Mirzoyan, R.}, {Mizuno, T.}, {Molero
  Gonzalez, M.}, {Molina, E.}, {Montaruli, T.}, {Monteiro, I.}, {Moralejo, A.},
  {Morcuende, D.}, {Morselli, A.}, {Mrakovcic, K.}, {Murase, K.}, {Nagai, A.},
  {Nakamori, T.}, {Nickel, L.}, {Nievas, M.}, {Nishijima, K.}, {Noda, K.},
  {Nosek, D.}, {Nozaki, S.}, {Ohishi, M.}, {Ohtani, Y.}, {Okazaki, N.},
  {Okumura, A.}, {Orito, R.}, {Otero-Santos, J.}, {Palatiello, M.}, {Paneque,
  D.}, {Pantaleo, F. R.}, {Paoletti, R.}, {Paredes, J. M.}, {Pavletić, L.},
  {Pech, M.}, {Pecimotika, M.}, {Pietropaolo, E.}, {Pirola, G.}, {Podobnik,
  F.}, {Poireau, V.}, {Polo, M.}, {Pons, E.}, {Prandini, E.}, {Prast, J.},
  {Priyadarshi, C.}, {Prouza, M.}, {Rando, R.}, {Rhode, W.}, {Ribó, M.},
  {Rizi, V.}, {Rodriguez Fernandez, G.}, {Saito, T.}, {Sakurai, S.}, {Sanchez,
  D. A.}, {Šarić, T.}, {Saturni, F. G.}, {Scherpenberg, J.}, {Schleicher,
  B.}, {Schmuckermaier, F.}, {Schubert, J. L.}, {Schussler, F.}, {Schweizer,
  T.}, {Seglar Arroyo, M.}, {Sitarek, J.}, {Sliusar, V.}, {Spolon, A.},
  {Strišković, J.}, {Strzys, M.}, {Suda, Y.}, {Sunada, Y.}, {Tajima, H.},
  {Takahashi, M.}, {Takahashi, H.}, {Takata, J.}, {Takeishi, R.}, {Tam, P. H.
  T.}, {Tanaka, S. J.}, {Tateishi, D.}, {Temnikov, P.}, {Terada, Y.},
  {Terauchi, K.}, {Terzic, T.}, {Teshima, M.}, {Tluczykont, M.}, {Tokanai, F.},
  {Torres, D. F.}, {Travnicek, P.}, {Truzzi, S.}, {Tutone, A.}, {Uhlrich, G.},
  {Vacula, M.}, {Vázquez Acosta, M.}, {Verguilov, V.}, {Viale, I.}, {Vigliano,
  A.}, {Vigorito, C. F.}, {Vitale, V.}, {Voutsinas, G.}, {Vovk, I.},
  {Vuillaume, T.}, {Walter, R.}, {Will, M.}, {Yamamoto, T.}, {Yamazaki, R.},
  {Yoshida, T.}, {Yoshikoshi, T.}, {Zywucka, N.}, {Balbo, M.}, {Eckert, D.}, \&
  {Tramacere, A.}}]{Abe2023_Multiwavelength}
{Abe}, S., {Aguasca-Cabot, A.}, {Agudo, I.}, {et~al.} 2023, A\&A, 673, A75

\bibitem[{Adams {et~al.}(2025)Adams, Bangale, Benbow, Buckley, Chen,
  Christiansen, Chromey, Escobar~Godoy, Feldman, Feng, Foote, Fortson, Furniss,
  Hanlon, Hervet, Hinrichs, Holder, Hughes, Humensky, Jin, Kaaret, Kertzman,
  Kherlakian, Kieda, Kleiner, Korzoun, Kumar, Lang, Lundy, Maier, Millard,
  Moriarty, Mukherjee, Ning, Ong, Pohl, Pueschel, Quinn, Rabinowitz, Ragan,
  Reynolds, Ribeiro, Roache, Sadeh, Saha, Sembroski, Shang, Tak, Talluri,
  Tucci, Valverde, Williams, Wong, Woo, collaboration), Alfaro, Alvarez,
  Arteaga-Velázquez, Avila~Rojas, Babu, Belmont-Moreno, Bernal,
  Caballero-Mora, Carramiñana, Casanova, Cotti, Cotzomi, De~la Fuente,
  de~León, Depaoli, Desiati, Di~Lalla, Diaz~Hernandez, DuVernois, Engel,
  Ergin, Espinoza, Fan, Fraija, Fraija, García-González, Garfias,
  Gonzalez~Muñoz, González, Goodman, Groetsch, Harding, Hernández-Cadena,
  Herzog, Huang, Hueyotl-Zahuantitla, Hüntemeyer, Iriarte, Kaufmann, Lara,
  Lee, León~Vargas, Longinotti, Luis-Raya, Malone, Martinez, Martínez-Castro,
  Matthews, Miranda-Romagnoli, Morales-Soto, Moreno, Araya, Mostafá, Najafi,
  Nayerhoda, Nellen, Omodei, Ponce, Pérez-Pérez, Rho, Rosa-González, Roth,
  Salazar, Sandoval, Schneider, Serna-Franco, Smith, Son, Springer, Tibolla,
  Tollefson, Torres, Torres-Escobedo, Turner, Ureña-Mena, Varela, Villaseñor,
  Wang, Wang, Watson, Wu, Yu, Yun-Cárcamo, Zhou, Martin, collaboration), Mori,
  Hailey, Safi-Harb, Zhang, \& collaboration)}]{Adams_2025}
Adams, C.~B., Bangale, P., Benbow, W., {et~al.} 2025, ApJ, 991, 192

\bibitem[{{Aharonian} \& {Atoyan}(1996)}]{Aharonian&Atoyan_1996}
{Aharonian}, F.~A. \& {Atoyan}, A.~M. 1996, \aap, 309, 917

\bibitem[{{Bordas} {et~al.}(2009){Bordas}, {Bosch-Ramon}, {Paredes}, \&
  {Perucho}}]{2009A&A...497..325B}
{Bordas}, P., {Bosch-Ramon}, V., {Paredes}, J.~M., \& {Perucho}, M. 2009, \aap,
  497, 325

\bibitem[{{Bosch-Ramon} {et~al.}(2005){Bosch-Ramon}, {Aharonian}, \&
  {Paredes}}]{valenti2005}
{Bosch-Ramon}, V., {Aharonian}, F.~A., \& {Paredes}, J.~M. 2005, A\&A, 432, 609

\bibitem[{Cao {et~al.}(2021)Cao, Aharonian, An, Axikegu, Bai, Bai, Bao,
  Bastieri, Bi, Bi, Cai, Cai, Cao, Chang, Chang, Chen, Chen, Chen, Chen, Chen,
  Chen, Chen, Chen, Chen, Chen, Chen, Chen, Chen, Cheng, Cheng, Cui, Cui, Cui,
  Piazzoli, Dai, Dai, Dai, Dan-Zeng-Luo-Bu, Volpe, Dong, Duan, Fan, Fan, Fan,
  Fang, Fang, Feng, Feng, Feng, Feng, Gao, Gao, Gao, Gao, Gao, Ge, Geng, Gong,
  Gou, Gu, Guo, Guo, Guo, Guo, Guo, Han, He, He, He, He, He, He, Heller, Hor,
  Hou, Hu, Hu, Hu, Hu, Huang, Huang, Huang, Huang, Huang, Huang, Ji, Ji, Jia,
  Jiang, Jiang, Jin, Ke, Kuleshov, Levochkin, Li, Li, Li, Li, Li, Li, Li, Li,
  Li, Li, Li, Li, Li, Li, Li, Li, Li, Liang, Liang, Lin, Liu, Liu, Liu, Liu,
  Liu, Liu, Liu, Liu, Liu, Liu, Liu, Liu, Liu, Liu, Liu, Liu, Long, Lu, Lv, Ma,
  Ma, Ma, Mao, Masood, Min, Mitthumsiri, Montaruli, Nan, Pang,
  Pattarakijwanich, Pei, Qi, Qi, Qiao, Qin, Ruffolo, Rulev, Sáiz, Shao,
  Shchegolev, Sheng, Shi, Song, Stenkin, Stepanov, Su, Sun, Sun, Sun, Tam,
  Tang, Tian, Wang, Wang, Wang, Wang, Wang, Wang, Wang, Wang, Wang, Wang, Wang,
  Wang, Wang, Wang, Wang, Wang, Wang, Wang, Wang, Wang, Wang, Wang, Wei, Wei,
  Wei, Wen, Wu, Wu, Wu, Wu, Wu, Xi, Xia, Xia, Xiang, Xiao, Xiao, Xiao, Xin,
  Xin, Xing, Xu, Xu, Xue, Yan, Yan, Yang, Yang, Yang, Yang, Yang, Yang, Yang,
  Yao, Yao, Ye, Yin, Yin, You, You, Yu, Yuan, Zeng, Zeng, Zeng, Zeng, Zha,
  Zhai, Zhang, Zhang, Zhang, Zhang, Zhang, Zhang, Zhang, Zhang, Zhang, Zhang,
  Zhang, Zhang, Zhang, Zhang, Zhang, Zhang, Zhang, Zhang, Zhang, Zhao, Zhao,
  Zhao, Zhao, Zhao, Zheng, Zheng, Zhou, Zhou, Zhou, Zhou, Zhou, Zhou, Zhu, Zhu,
  Zhu, Zhu, \& Zuo}]{Cao2021_Discovery}
Cao, Z., Aharonian, F., An, Q., {et~al.} 2021, \apjl, 919, L22

\bibitem[{Cao {et~al.}(2024)Cao, Aharonian, An, Axikegu, Bai, Bao, Bastieri,
  Bi, Bi, Cai, Cao, Cao, Cao, Chang, Chang, Chen, Chen, Chen, Chen, Chen, Chen,
  Chen, Chen, Chen, Chen, Chen, Chen, Cheng, Cheng, Cui, Cui, Cui, Cui, Dai,
  Dai, Dai, Danzengluobu, della Volpe, Dong, Duan, Fan, Fan, Fang, Fang, Feng,
  Feng, Feng, Feng, Feng, Gabici, Gao, Gao, Gao, Gao, Gao, Gao, Ge, Geng,
  Giacinti, Gong, Gou, Gu, Guo, Guo, Guo, Guo, Han, He, He, He, He, He, Heller,
  Hor, Hou, Hou, Hou, Hu, Hu, Hu, Huang, Huang, Huang, Huang, Huang, Huang,
  Huang, Ji, Jia, Jia, Jiang, Jiang, Jiang, Jin, Kang, Ke, Kuleshov, Kurinov,
  Li, Li, Li, Li, Li, Li, Li, Li, Li, Li, Li, Li, Li, Li, Li, Li, Li, Li, Li,
  Liang, Liang, Lin, Liu, Liu, Liu, Liu, Liu, Liu, Liu, Liu, Liu, Liu, Liu,
  Liu, Liu, Liu, Lu, Luo, Lv, Ma, Ma, Ma, Mao, Min, Mitthumsiri, Mu, Nan,
  Neronov, Ou, Pang, Pattarakijwanich, Pei, Qi, Qi, Qiao, Qin, Ruffolo, Sáiz,
  Semikoz, Shao, Shao, Shchegolev, Sheng, Shu, Song, Stenkin, Stepanov, Su,
  Sun, Sun, Sun, Tam, Tang, Tang, Tian, Wang, Wang, Wang, Wang, Wang, Wang,
  Wang, Wang, Wang, Wang, Wang, Wang, Wang, Wang, Wang, Wang, Wang, Wang, Wang,
  Wang, Wang, Wei, Wei, Wei, Wen, Wu, Wu, Wu, Wu, Wu, Xi, Xia, Xia, Xiang,
  Xiao, Xiao, Xin, Xin, Xing, Xiong, Xu, Xu, Xu, Xu, Xue, Yan, Yan, Yan, Yang,
  Yang, Yang, Yang, Yang, Yang, Yang, Yang, Yang, Yao, Yao, Ye, Yin, Yin, You,
  You, Yu, Yuan, Yue, Zeng, Zeng, Zeng, Zha, Zhang, Zhang, Zhang, Zhang, Zhang,
  Zhang, Zhang, Zhang, Zhang, Zhang, Zhang, Zhang, Zhang, Zhang, Zhang, Zhang,
  Zhang, Zhang, Zhao, Zhao, Zhao, Zhao, Zhao, Zheng, Zhou, Zhou, Zhou, Zhou,
  Zhou, Zhou, Zhou, Zhu, Zhu, Zhu, Zhu, Zuo, \&
  Collaboration)}]{Cao2024_Catalog}
Cao, Z., Aharonian, F., An, Q., {et~al.} 2024, ApJSS, 271, 25

\bibitem[{Condon {et~al.}(1998)Condon, Cotton, Greisen, Yin, Perley, Taylor, \&
  Broderick}]{Condon_1998}
Condon, J.~J., Cotton, W.~D., Greisen, E.~W., {et~al.} 1998, \aj, 115, 1693

\bibitem[{{\noopsort{Dario}}{de la Fuente}
  {et~al.}({2023\natexlab{b}}){\noopsort{Dario}}{de la Fuente},
  {Toledano-Juárez, I.}, {Kawata, K.}, {Trinidad, M. A.}, {Yamagishi, M.},
  {Takekawa, S.}, {Tafoya, D.}, {Ohnishi, M.}, {Nishimura, A.}, {Kato, S.},
  {Sako, T.}, {Takita, M.}, {Sano, H.}, \& {Yadav, R. K.}}]{deLaFuente2023_MC2}
{\noopsort{Dario}}{de la Fuente}, E., {Toledano-Juárez, I.}, {Kawata, K.},
  {et~al.} {2023\natexlab{b}}, A\&A, 675, L5

\bibitem[{de~la Fuente {et~al.}(2023{\natexlab{a}})de~la Fuente,
  Toledano-Juarez, Kawata, Trinidad, Tafoya, Sano, Tokuda, Nishimura, Onishi,
  Sako, Hona, Ohnishi, \& Takita}]{deLaFuente2023_MC}
de~la Fuente, E., Toledano-Juarez, I., Kawata, K., {et~al.} 2023{\natexlab{a}},
  PASJ, 75, 546

\bibitem[{De~Sarkar(2023)}]{DeSarkar2023_SNRModel}
De~Sarkar, A. 2023, MNRAS: Letters, 521, L5

\bibitem[{Fujita {et~al.}(2010)Fujita, Ohira, \& Takahara}]{Fujita_2010}
Fujita, Y., Ohira, Y., \& Takahara, F. 2010, \apjl, 712, L153

\bibitem[{Gabici {et~al.}(2007)Gabici, Aharonian, \& Blasi}]{Gabici+2007}
Gabici, S., Aharonian, F.~A., \& Blasi, P. 2007, \apss, 309, 365

\bibitem[{{Kaiser} \& {Alexander}(1997)}]{1997MNRAS.286..215K}
{Kaiser}, C.~R. \& {Alexander}, P. 1997, \mnras, 286, 215

\bibitem[{Kelner {et~al.}(2006)Kelner, Aharonian, \& Bugayov}]{Kelner2006}
Kelner, S.~R., Aharonian, F.~A., \& Bugayov, V.~V. 2006, Phys. Rev. D, 74,
  034018

\bibitem[{{Mahanta} {et~al.}(2025){Mahanta}, Fazio, Hora, Allen, Ashby, Barmby,
  Deutsch, Huang, Kleiner, Marengo, Megeath, Melnick, Pahre, Patten, Polizotti,
  Smith, Taylor, Wang, Willner, Hoffmann, Pipher, Forrest, McMurty, McCreight,
  McKelvey, \& McMurray}]{Mahanta2024_uGMRT}
{Mahanta}, G.~K., Fazio, G.~G., Hora, J.~L., {et~al.} 2025, JHEAP, 47, 100381

\bibitem[{{Mart{\'\i}} {et~al.}(2026){Mart{\'\i}}, {Luque-Escamilla},
  {Paredes}, \& {Mart{\'\i}nez Aroza}}]{Marti2026_LHAASO}
{Mart{\'\i}}, J., {Luque-Escamilla}, P.~L., {Paredes}, J.~M., \& {Mart{\'\i}nez
  Aroza}, J. 2026, \mnras, 546, stag297

\bibitem[{{Mitchell}(2024)}]{Mitchell2024_SNRCloud}
{Mitchell}, A. M.~W. 2024, A\&A, 684, A66

\bibitem[{{Miville-Desch{\^e}nes} {et~al.}(2017){Miville-Desch{\^e}nes},
  {Murray}, \& {Lee}}]{Miville2017}
{Miville-Desch{\^e}nes}, M.-A., {Murray}, N., \& {Lee}, E.~J. 2017, \apj, 834,
  57

\bibitem[{Nava \& Gabici(2012)}]{NavaGabici2012}
Nava, L. \& Gabici, S. 2012, \mnras, 429, 1643

\bibitem[{{Spitzer}(1978)}]{Spitzer-Book1978}
{Spitzer}, L. 1978, {Physical processes in the interstellar medium. A
  Wiley-Interscience Publication, New York: Wiley.}

\bibitem[{{Wang} {et~al.}(2025){Wang}, {Reville}, \& {Aharonian}}]{Wang2025ApJ}
{Wang}, J., {Reville}, B., \& {Aharonian}, F.~A. 2025, \apjl, 989, L25

\end{thebibliography}

\begin{appendix}

\section{Details of the physical model}
\label{sec:model_app}

\begin{table}
\begin{center}
\caption{Parameters of the model. }
\label{tab:parametros generales}
\begin{adjustbox}{max width=\columnwidth}
\begin{tabular}{l c c c}
\hline
\hline
\rule{0pt}{2.5ex}Parameter & Value & Units  \\
\hline
\rule{0pt}{2.5ex}Age of the MQ [$t_0$]  & $10^4$ & ${\rm yr}$ \\
Distance to the system$^\dagger$ [$d$]  & $3.3$ & ${\rm kpc}$\\
Jet mechanical power [$L_{\rm j}$]   & $3\times 10^{39}$ & ${\rm erg\,{s}^{-1}}$\\
Fraction of power to relativistic particles [$q_{\rm rel}$]   & $0.1$ & \\
Hadron-to-lepton power ratio  & $100$ & \\
Spectral index [$p$]   & $2$ & \\
Density of the ISM [$n_{\rm ISM}$]   & $0.1$ & ${\rm cm^{-3}}$ \\
Diffusion coefficient at 10 GeV [$D_{\rm 0}$]  & $10^{26}$ & ${\rm cm^2 \, s^{-1}}$\\
Cocoon semi-major axis at $t_0$ [$l_{\rm c}$] & $20$ & ${\rm pc}$ \\
Cocoon number density & $10^{-3}$ & ${\rm cm^{-3}}$\\
Cocoon magnetic field & $3$ & ${\mu \rm G}$\\
Cocoon radiation energy density (CMB) & $0.26$ & ${\rm eV\,cm^{-3}}$\\
Number density for both effective regions$^\dagger$ [$n$] & $30$ & ${\rm cm^{-3}}$\\
Effective distance to $C_1$ [$R_1$]  & $1$ & ${\rm pc}$\\
Radius of $C_1$ [$r_1$]  & $1$ & ${\rm pc}$\\
Effective distance to $C_2$ [$R_2$]  & $35$ & ${\rm pc}$\\
Radius of $C_2$ [$r_2$]  & $12$ & ${\rm pc}$\\
\hline
\end{tabular}

\end{adjustbox}
{\footnotesize $^\dagger$ Value corresponding to the molecular cloud  MML[2017]4607 \citep[see][]{Miville2017,deLaFuente2023_MC}.}
\tablefoot{We assume that all parameters are constant over time, except for the cocoon length ($l_{\rm c}\equiv l_{\rm c}(t)$).}

\end{center}
\end{table}

\begin{figure}
    \centering
    \includegraphics[width=\columnwidth]{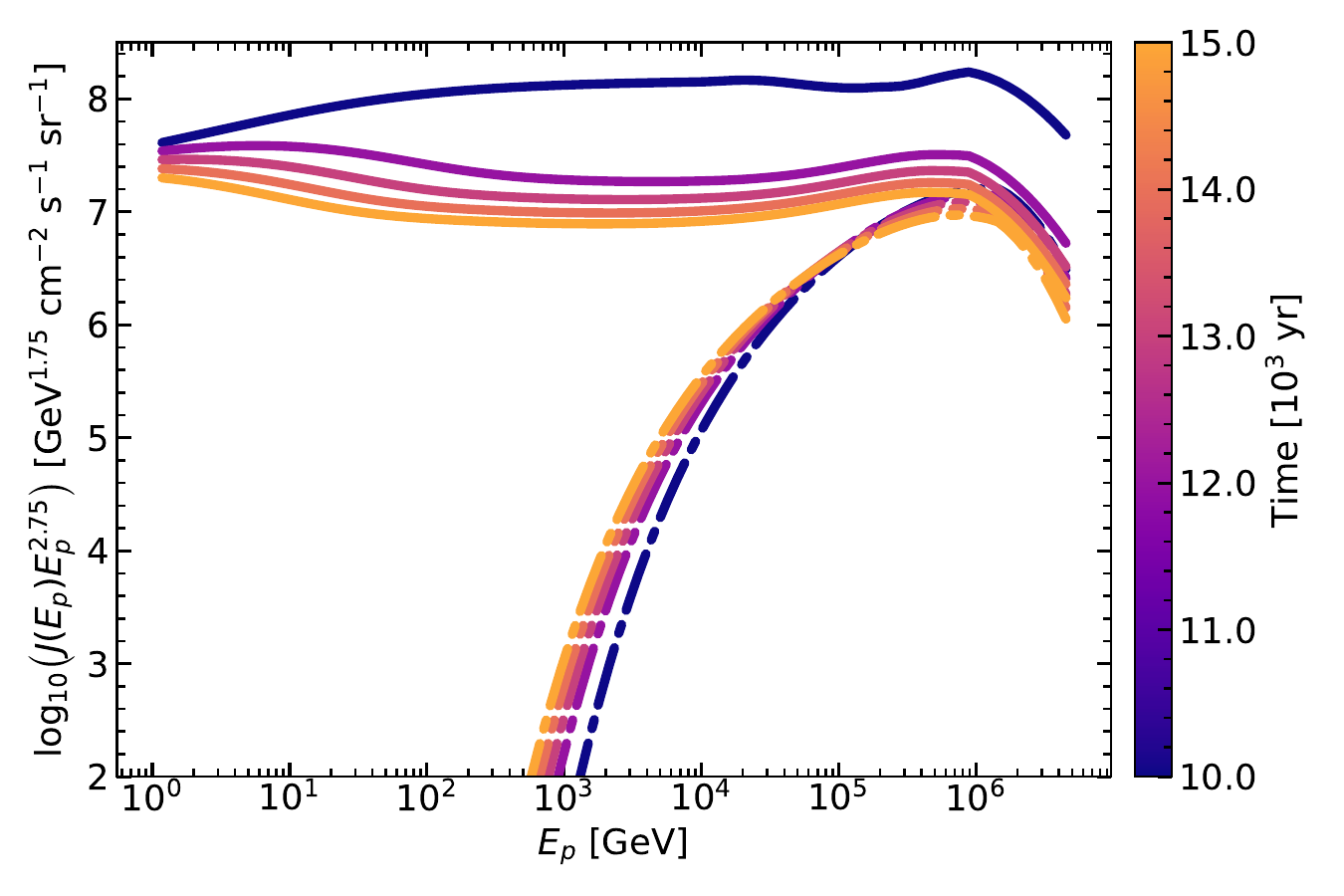}
    \caption{Distribution of protons propagated through the ISM. Solid lines correspond to protons reaching $C_1$ while dotted-dashed lines show the same for $C_2$. The colour bar indicates the age of the system, $t_{\rm total}$.}
    \label{fig: p_distribution}
\end{figure}

Table \ref{tab:parametros generales} lists the main parameters of our model. In the next section we explore the parameter sensitivity of the model. 

Regarding particle distribution and transport, after injection, protons propagate in the ISM under energy-dependent
diffusion. Energy losses for
protons are dominated by inelastic $pp$ interactions in dense
regions.
As discussed in Sect. 4, the target material is represented by two effective interaction regions
within the MML[2017]4607 cloud at \(d=3.3\) kpc, both with a characteristic
density \(n\simeq30\,{\rm cm^{-3}}\).

With a physical radius $r_1 = 1$ pc, this inner effective region intercepts the confined GeV population. The second is a distant effective region, $C_2$, situated at $R_2 \sim 35$ pc with a radius $r_2 = 12$ pc, acting as the primary target for the fast-escaping PeVatron component.
A more refined treatment including a spatially extended source instead of a point-like injection site would not alter the qualitative transport-driven interpretation, although it could slightly affect the normalisation of the nearby component.

The proton density at each effective target region is obtained by
convolving the time-dependent escape history from the cocoon with the
ISM diffusion Green function, as described in Sect.~3. After jet switch-off,
no fresh particles are injected by the engine, but particles stored in the
cocoon continue to escape and propagate through the surrounding medium. The resulting particle densities at each location are
used to compute the $\gamma$-ray emissivity via inelastic $pp$
interactions following standard prescriptions.

Figure~\ref{fig: p_distribution} shows the evolution of the propagated
proton distribution evaluated at the locations of the two representative
effective interaction regions, $C_1$ ($R_1=1$ pc; solid lines) and
$C_2$ ($R_2=35$ pc; dot--dashed lines). The colour-scale indicates the
total system age, $t_{\rm total}$.

For the inner effective region \(C_1\), the diffusion time over the
effective scale \(R_1\) is negligible (\(t_{\rm diff}\ll1\) kyr), meaning
that the interaction of the less energetic GeV hadrons with nearby dense
material can proceed over essentially the full age of the system. These
particles remain largely confined to the inner region, and its emission is
therefore represented by the late-time state (\(t\simeq13\) kyr, violet curve). In contrast, the more distant effective region $C_2$ ($35$~pc) accumulates PeV protons during the lifetime of the MQ. Therefore, while $C_1$ currently intercepts the slow-diffusing GeV population left behind, the high-energy emission at $C_2$ naturally reflects the fast-escaping PeVatron component.

As may be seen,  the proton
spectrum at a fixed location is not stationary but evolves with time. This time synchronisation naturally explains the spectral dichotomy between $C_1$ and $C_2$: $C_2$ samples the high-energy, less softened part of
the original injected proton population, while $C_1$ captures the accumulated,
softer GeV population.   
This scenario lends physical plausibility to the interpretation of LHAASO~J2108+5157 as a fossil MQ observed at an intermediate epoch, consistent with expectations for MQRs under energy-dependent diffusion \citep{Abaroa_etal_2026MQR}.

Primary electrons injected at the jet termination region are
treated consistently within the same framework. Their evolution
includes synchrotron and inverse Compton losses in a magnetic
field $B=3\,\mu$G and radiation fields dominated by the cosmic
microwave background and the local Galactic interstellar radiation field. The corresponding synchrotron and inverse
Compton emission from the cocoon is calculated to assess the
detectability of the relic structure at radio and X-ray
frequencies.
While low-energy radio-emitting electrons may survive the radiative cooling, the 
diffuse radio emission falls below current survey sensitivities over such an extensive area in the cocoon. Assuming a reasonable physical diameter
 of $\sim 50$~pc (corresponding to an angular scale of $\sim 0.86^\circ$ subtending $\simeq 0.58$ deg$^2$ at $3.3\,{\rm kpc}$) and 
using the NRAO VLA Sky Survey \citep[NVSS;][]{Condon_1998} at 1.4~GHz ($\sigma_{\rm beam} \approx 0.45$~mJy~beam$^{-1}$; $45^{\prime\prime}$ resolution), we can integrate the noise in quadrature over this $0.58$~deg$^2$ region ($N \approx 3360$ beams). This yields a $3\sigma$ upper limit of $S_{\nu} \approx 77$~mJy, 
corresponding to an energy flux of $\nu F_\nu \approx 1 \times 10^{-14}~\mathrm{erg~cm^{-2}~s^{-1}}$, which we incorporate into our broadband modelling.

All model parameters (see Table \ref{tab:parametros generales}) are chosen within ranges typical of powerful
high-mass MQs.

\section{Sensitivity tests}

To assess the sensitivity of our results to the main transport
parameters, 
we computed three additional models in which one
parameter is varied at a time with respect to the fiducial case
presented in the main text. This exercise is not intended as a complete multidimensional fit but rather as an illustrative check of
how the transport-driven separation between the GeV and TeV--PeV components responds
to moderate changes in the assumed physical conditions. The parameters explored are the
normalisation of the diffusion coefficient, $D_0$, the energy
dependence of diffusion, $\delta$, and the duration of the active
MQ phase, $t_{\rm MQ}$. These are the quantities that most
directly control the time at which particles of a given energy
can reach the distant molecular material.

\begin{figure}
  \centering
    \includegraphics[width=\columnwidth]{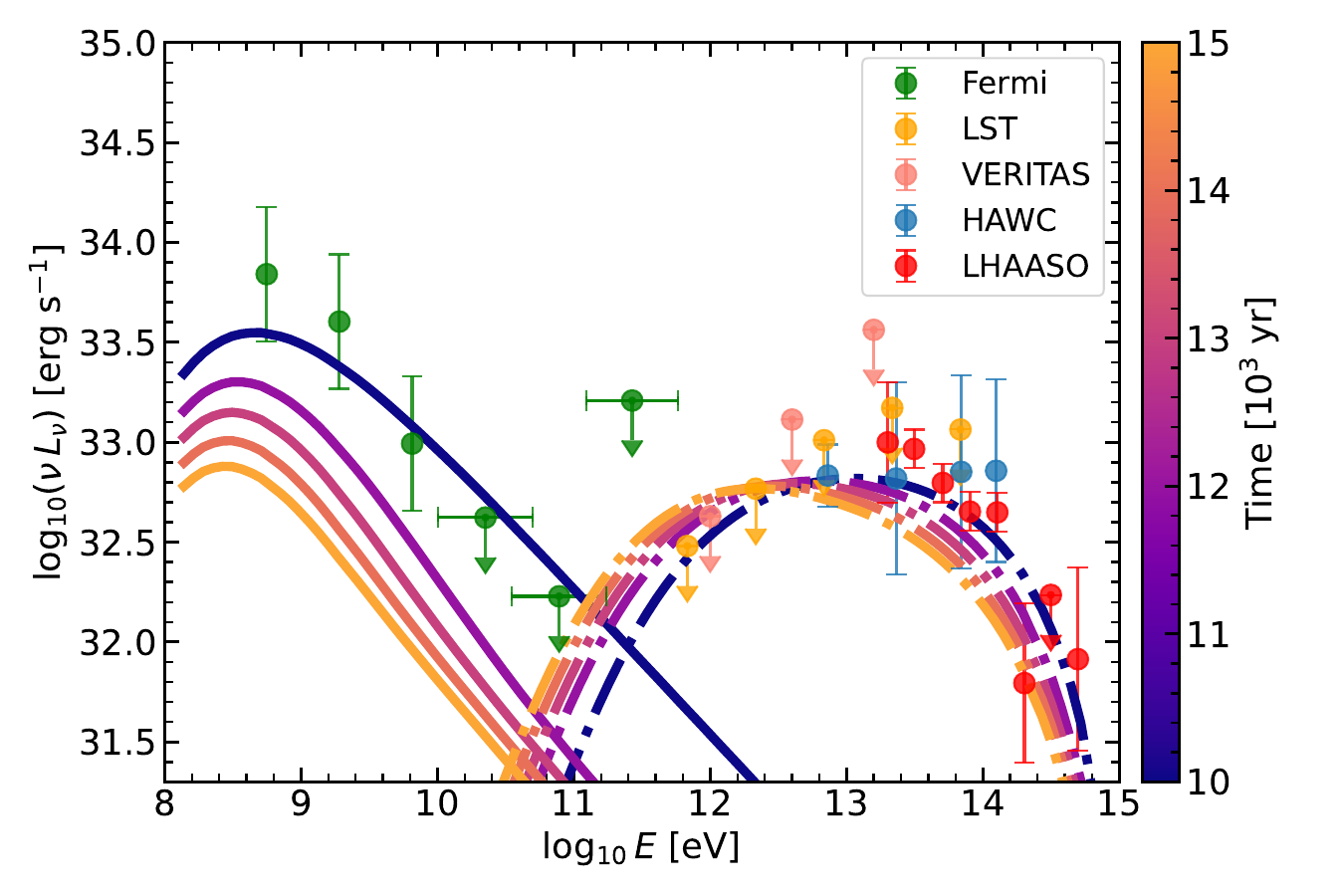} 
    \includegraphics[width=\columnwidth]{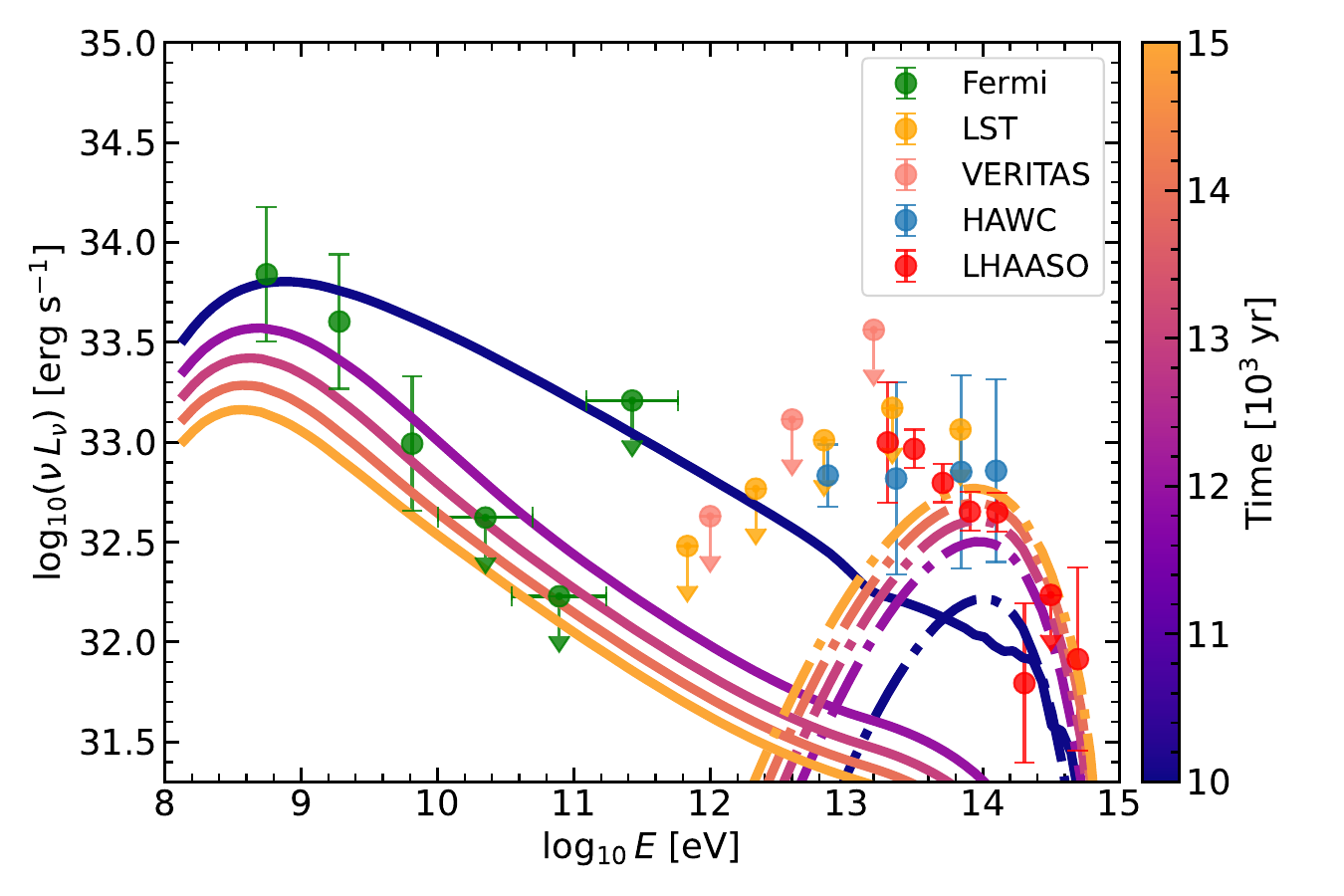}
    \includegraphics[width=\columnwidth]{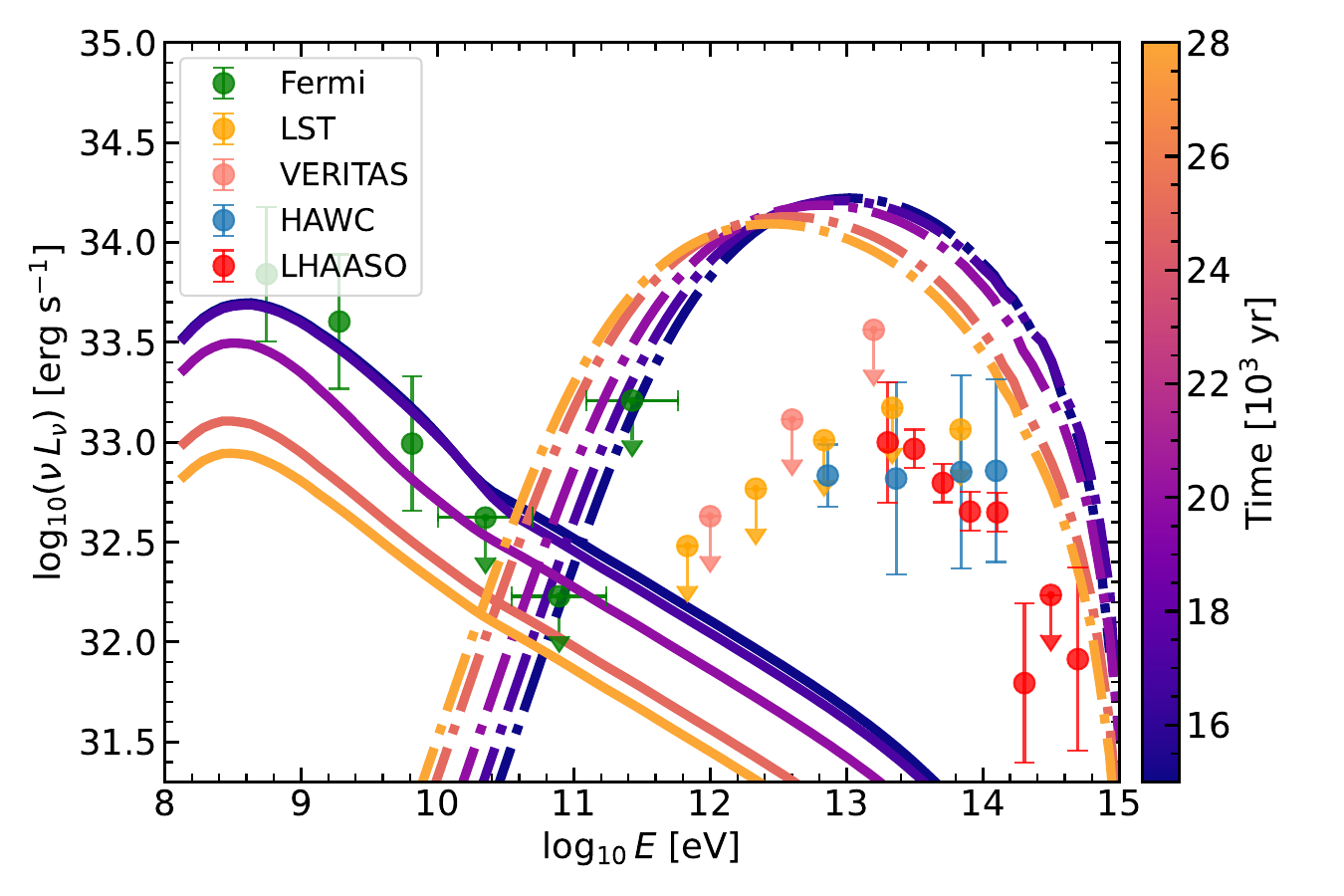}
  \caption{Sensitivity of the SEDs to selected transport parameters. 
The panels show models in which one parameter is varied with respect 
to the fiducial case shown in Fig.~\ref{fig:sed}: $D_0=2\times10^{26}\,{\rm cm^2\,s^{-1}}$ 
(\textit{top}), $\delta=0.3$ (\textit{middle}), and $t_{\rm MQ}=15$ kyr (\textit{bottom}). 
Solid and dot-dashed lines correspond to the emission from $C_1$ and $C_2$, 
respectively. Colours indicate the total system age. The comparison illustrates the sensitivity of the GeV/TeV--PeV separation to the
transport parameters, rather than providing a full multidimensional scan.}
  \label{fig: alternative_models}
\end{figure}

Figure~\ref{fig: alternative_models} shows the resulting SEDs.
In the top panel, we adopt $D_0=2\times10^{26}\,
{\rm cm^2\,s^{-1}}$, twice the fiducial value. A larger diffusion
coefficient allows particles to reach the distant interaction region more
rapidly, shifting the contribution from $C_2$ to earlier evolutionary
times and modifying the depth and location of the spectral
valley between the GeV and TeV bands. In the middle panel,
we keep the diffusion normalisation fixed but use a weaker
energy dependence, $\delta=0.3$. In this case, the contrast
between the propagation of low- and high-energy particles is
reduced, and the transport-induced segregation between $C_1$ and
$C_2$ becomes less efficient. Finally, in the bottom panel we assume
a longer active MQ phase, $t_{\rm MQ}=15$ kyr. This increases the
duration of particle injection and modifies the escape history
from the cocoon, producing a broader and brighter high-energy
component than in the fiducial model.

These tests should be interpreted as illustrative sensitivity checks, not as a full
exploration of the multidimensional parameter space. They show that the relative
location and normalisation of the GeV and TeV--PeV components are sensitive to the
transport history, in particular to the diffusion coefficient, its energy dependence,
and the source age. Moderate changes in these parameters shift the timing and
spectral location of the two interaction regions and can therefore worsen the agreement
with the observed spectral valley.

At the same time, degeneracies remain. The relic age,
\(t_{\rm MQR}\), is partly sampled by the different evolutionary times
shown in Figs.~1 and B.1, but we do not perform an independent
optimisation over \(t_{\rm MQ}\) and \(t_{\rm MQR}\). Likewise, correlated
changes in distance, diffusion coefficient, source offset, target geometry,
and gas density could lead to alternative viable realisations. In particular,
changes in distance, source offset, or the detailed \(C_1/C_2\) geometry
would rescale the physical separations and diffusion times and would
therefore require a dedicated re-optimisation of the model.

The density and size of the interaction regions were not varied explicitly in Fig.~B.1 because, in the optically thin \(pp\) regime considered here,
they mainly rescale the hadronic luminosity through
\(L_\gamma \propto n_{\rm H}V_{\rm reg}\), while leaving the transport-induced spectral separation largely unchanged. The hadron-to-lepton ratio controls the relative strength of the hadronic and
leptonic channels, while the injection cutoff affects mostly the highest-energy tail of the \(C_2\) component. 

Thus, the fiducial parameters should not be interpreted as a unique
best-fit solution. Rather, they identify the approximate physical regime
required by the MQR scenario: slow diffusion, kyr-scale relic evolution,
and a hadronic energy budget compatible with powerful MQ jets. A full
morphological and multidimensional fit is beyond the scope of this work.

\end{appendix}

\end{document}